\documentclass[a4paper,10pt]{revtex4}
\usepackage{mathrsfs}
\usepackage{graphicx}
\usepackage{latexsym}
\usepackage{amsmath}
\usepackage{amssymb}
\usepackage{textcomp}
\usepackage{amsbsy}
\usepackage{graphics}
\usepackage{epstopdf}

\begin{document}

\title{Graviton Kaluza Klein modes in non-flat branes with stabilised modulus}
\author{Tanmoy Paul}
\email{tpap@iacs.res.in}
\author{Soumitra SenGupta}
\email{tpssg@iacs.res.in}
\affiliation{Department of Theoretical Physics,\\
Indian Association for the Cultivation of Science,\\
2A $\&$ 2B Raja S.C. Mullick Road,\\
Kolkata - 700 032, India.\\}

\begin{abstract}
We consider a generalised two brane Randall Sundrum model where the branes are endowed with non-zero cosmological constant.
In this scenario, we re-examine the modulus stabilisation mechanism and the nature of Kaluza Klein (KK) graviton modes. 
 Our result reveals that while the KK mode 
graviton masses may change significantly with brane cosmological constant, the Goldberger-Wise stabilisation mechanism, which assumes negligible  
backreaction on the background metric, continues to hold even when the branes have large cosmological constant. 
The possibility of having a global minimum for the modulus is also discussed.
\end{abstract}

\maketitle
\section{Introduction}
Gauge hierarchy problem continues to be an unsolved issue in the standard model of elementary particles despite its 
enormous  success in describing Physics up to TeV scale. A solution to the gauge hierarchy problem was 
proposed by Randall and Sundrum (RS model) by considering an extra dimension compactified into a circle $S^1$ with a 
$Z_2$ orbifolding \cite{RS}. The modulus corresponding to the radius of the extra dimension in such a model can be 
stabilised via Goldberger-Wise (GW) stabilization mechanism \cite{GW}. Both RS and GW model 
do not invoke any intermediate scale in the theory and are robust against radiative corrections.
This resulted into a large volume of work   in particle phenomenology and cosmology 
in the backdrop of Randall-Sundrum warped geometry scenario. In the context of collider Physics, 
possible role of first Kaluza-Klein (KK) graviton mode have been extensively studied  in ATLAS and CMS detectors at 
LHC which have already set a stringent lower bound for the mass of the first KK graviton to be $m_{n=1} \sim 2.5$TeV \cite{atlas1,atlas2}. Various
implications of this have been discussed in \cite{dhr,ashmita,rizzo,ashmita1,yong,dhr1,thomas}. 

There have been several efforts to formulate some variants of the original RS model. One such effort addresses a similar 
warped geometry model with non flat 3-branes in contrast to the original RS model which assumes two flat 3-branes sitting 
at the two orbifold fixed points. It has been shown in \cite{ssg} that one can indeed generalise the model with non-zero cosmological 
constant on the visible 3-brane i.e on our observable universe and can resolve the gauge hierarchy problem concomitantly. In this generalised 
RS model \cite{joydip,koley} it has been shown that the 3-branes can be either de-Sitter (dS) or anti de-Sitter (AdS) where the magnitude of the 
induced cosmological constant and 
that of the warping parameter are intimately connected. It is therefore 
crucially important to determine whether in such non-flat warped geometry models, the Goldberger-Wise stabilisation mechanism, which 
neglects the backreaction of the stabilising field, 
can still be employed successfully to stabilise the radius of the extra dimension to the desired 
value $\sim M_{Pl}^{-1}$. Moreover it is worth while to explore 
the effect of the brane cosmological 
constant on the masses of the graviton KK modes which are expected to play important role in high energy scattering processes. 
This work is focussed into addressing the 
following two questions in a generalised RS model:
\begin{enumerate}
 \item Can the modulus of extra dimension be stabilised to a global minimum for the entire range of values of the cosmological constant  
 in the context of generalised RS model?\\
 \item What are the KK graviton masses for different choices of cosmological constant?
\end{enumerate}
After a brief review of original and generalised RS model in first two sections, we focus into the modulus stabilization 
conditions as well as the expressions for the modified KK graviton masses due to the presence of non-vanishing brane cosmolgical 
constant.
\section{Randall Sundrum model}
In the RS scenario, it is predicted that there exists an extra spatial dimension in addition to the ( 3 + 1 ) dimensional observed universe.  The corresponding five dimensional bulk space-time  is described by a  metric
\begin{equation}
 ds^2=\exp{(-2k|y|)}\eta_{\mu\nu}dx^{\mu}dx^{\nu} -dy^2
 \label{originalR-Smetric}
\end{equation}
where greek indices $\mu$, $\nu$ run over 0,1,2,3 and refer to the four observed dimensions. The geometry of the extra dimension is
 $S^{1}/Z_2$  and is described by the coordinate 'y' . Here the circle $S^{1}$ has radius $r$. $\Lambda$ is the bulk cosmolgical 
 constant, $k=\sqrt{-\Lambda/12M^3}$. The factor $\exp{(-2k|y|)}$ is known as the warp factor. The $y =$ constant slices 
 at $y=0$ and at $y=\pi r$ are known as the hidden and the visible branes, the observable universe being identified with the 
 latter which has a negative brane tension as opposed to the hidden brane with a positive brane tension. It can be shown 
 that a mass parameter $m_0$ of the order of Planck scale is warped to a value TeV on the visible brane following the relation 
 $m=m_0\exp{(-kr\pi)}$, for $kr\sim11.6$. Thus in this picture, the stability of higgs mass 
 against large radiative correction is ensured by the warped geometry of the five dimensional spacetime.
 In this context the KK graviton mass modes are determined by 
 considering a small fluctuation around the flat metric with its KK decomposition . Some of these modes have masses $m_{n=0}=0$;
 $m_{n=1}=0.383$TeV; $m_{n=2}=0.702$TeV for $k/M_{pl}\sim0.1$ \cite{dhr}. The requirement of $k<M_{pl}$ emerges from the fact that 
 k, which measures the bulk curvature must be smaller than the Planck scale so that the classical solutions of Einstein's equations 
 in the bulk  can be trusted \cite{dhr}.
 
 In the context of modulus stabilisation, it was proposed by Goldberger and Wise 
 that the modulus of the model (i.e. the radius of the extra dimension) can be stabilised to the desired value by introducing a massive
 scalar field in the bulk. Evaluating  the effective modulus potential due to the massive scalar field of mass m, one gets the stabilisation condition 
 as $kr=m^2/{(4\pi k^2)}\ln{(v_h/v_v)}$ where $v_h/v_v$ is the  ratio of the vacuum expectation values ( vev) of 
 the scalar field on hidden and visible brane. 
Taking $v_h/v_v\sim1.45$ and $m/k\sim0.2$ one gets $kr\sim11.6$.
In this analysis it was further shown that both $v_h$ and $v_v$ ( in Planckian unit ) must 
be smaller than unity so that the effect of back-reaction on the background metric can be ignored. 
Moreover the condition of having a global minimum for the modulus potential was found to be $\delta V_v<kv_v^2$, where $\delta V_v$
is a perturbation on the visible brane tension.  
\section{Generalised Randall Sundrum Model}
Present cosmological observation indicates the possible existence of a 4-dimensional cosmological constant 
($\sim10^{-124}$) in Planckian unit. It has been demonstrated \cite{ssg} that by relaxing the condition of zero cosmological 
constant (i.e flat 3-brane) it is possible to obtain a more general expression for the warp factor. 
Starting from a general metric ansatz, 
\begin{equation}
 ds^2=\exp{[-2A(y)]}g_{\mu\nu}dx^{\mu}dx^{\nu} -dy^2
 \label{generalisedR-Smetric}
\end{equation}
one may solve the bulk equations for both anti de-Sitter(AdS) and de-Sitter (dS) 3-branes. The corresponding warp factor 
for AdS brane is 
\begin{equation}
 \exp{[-A(y)]}=\omega \cosh{[\ln{(\omega/c_1)} + ky]}
 \label{AdSwarpfactor}\\
\end{equation}
with $ c_1=1 + \sqrt{1-\omega^2}$ and $\omega^2=-\varOmega/(3k^2)$ while that for dS brane is 
\begin{equation}
 \exp{[-A(y)]}=\omega \sinh{[\ln{(c_2/\omega)} - ky]}
  \label{dSwarpfactor}\\
\end{equation}
with $c_2=1 + \sqrt{1+\omega^2}$ and $\omega^2=\varOmega/(3k^2)$. Here $\varOmega$ is the brane cosmological constant and 
$\omega^2$ is a dimensionless parameter.
Just as in the original  RS model, this generalised scenario also can address the  gauge hierarchy problem  for appropriate choices of the parameters which we discuss below.

The scalar mass on the visible brane \cite{rubakov} gets warped through the warp factor. In order to resolve the gauge 
hierarchy problem it must satisfy, $\exp{[-A(\pi r)]}= 10^{-16}= m/m_0$. This leads to
\begin{equation}
 \exp{[-k\pi r]}=(10^{-16}/c_1)[1+\sqrt{1- \omega^2 10^{32}}]
 \label{AdSradius}
 \end{equation}
 for AdS case.
 \begin{equation}
 \exp{[-k\pi r]}=(10^{-16}/c_2)[1+\sqrt{1+ \omega^2 10^{32}}]
 \label{dSradius}
\end{equation}
for dS case.

From the above two relations one can say: $(1)$ real solution of $k\pi r$ exists which resolves the hierarchy problem,  
$(2)$ the the warping parameter  $kr$ depends on cosmological constant.\\
In the following section we  employ the GW stabilisation mechanism for the generalised RS model with non-flat branes to derive the new stability condition.
\section{Modulus stabilisation for non-flat branes }
To stabilise the modulus $r$ in the context of generalised RS model, we adopt the
method proposed by Goldberger and Wise \cite{GW}. Let us consider a massive scalar field $\Phi$
in the bulk with quartic interactions on the Planck ($y=0$) and visible branes ($y=\pi r$). 
The corresponding action is,
 \begin{eqnarray}
  S_5&=&(1/2)\int{ d^4xdy \sqrt{-g_5}[(\partial_{y}\Phi)^2 + m^2\Phi^2]}
    -\int d^4xdy \sqrt{-g_h}\lambda_h(\Phi^2-v_h^2)^2\delta{(y)}\nonumber\\
    &&-\int d^4xdy \sqrt{-g_v}\lambda_v(\Phi^2-v_v^2)^2\delta{(y-r\pi)}
  \label{scalaraction}
 \end{eqnarray}
 Here we assume that the scalar field depends only on extra dimensional coordinate. $g_h$ and
 $g_v$ are the determinants of the induced metric on the hidden and visible brane respectively.
 The vacuum expectation value of the scalar field on the branes are given by $v_h$ and $v_v$; $\lambda_h$ and $\lambda_v$ are brane tensions.
 
 The equation of motion for the scalar field is given by,
 \begin{equation}
  \Phi''(y)-4A'(y)\Phi'(y)-m^2\Phi+4\lambda_v(\Phi^2-v_v^2)\Phi \delta{(y-r\pi)}+4\lambda_h(\Phi^2-v_h^2)\Phi \delta{(y)}=0
  \label{scalarfieldequation}
 \end{equation}
 For large $\lambda_h$ and $\lambda_v$, one obtains the following two boundary conditions :
 \begin{eqnarray}
  \Phi{(0)}=v_h
  \label{boundarycondition1}\\
  \Phi{(\pi r)}=v_v
  \label{boundarycondition2}
 \end{eqnarray}
 Now we discuss the stability mechanism for  two different scenarios i.e. AdS and dS branes separately.
 \subsubsection{Anti de-Sitter brane ($\varOmega<0$)}
 It has been shown in \cite{ssg} that the magnitude of the cosmological constant on AdS brane is constrained to have an 
 upper bound and must lies between $-10^{-32}< \varOmega <0$.
 Due to this tiny value of the magnitude of the cosmological constant, we keep terms
 up to $\omega^2$ order. Differentiation of both sides of eqn.(\ref{AdSwarpfactor}) with respect to  'y' yields,
 \begin{equation}
  A'(y)=k[1-(\omega^2/2)\exp{(2ky)}]
  \nonumber
 \end{equation}
 Putting the above expression in eqn. (\ref{scalarfieldequation}), one gets the equation of motion 
 for scalar field in the bulk as,
 \begin{equation}
  \Phi''(y)-4k[1-(\omega^2/2)\exp{(2ky)}]\Phi'(y)-m^2\Phi=0
  \nonumber
 \end{equation}
 This leads to the solution,
 \begin{eqnarray}
 \Phi(y)&=&[A\exp{((2+\nu) ky)} + B\exp{((2-\nu) ky)}]
  - \omega^2/2[A(2+\nu)/(1+\nu)\exp{((4+\nu) ky)}\nonumber\\
 &&+B(2-\nu)/(1-\nu)\exp{((4-\nu) ky)}]
  \label{scalarfieldsolution}
 \end{eqnarray}
 Here A, B are arbitrary constants and $\nu =\sqrt{(4+m^2/k^2)}$. An effective potential $V_{eff}$
 can be obtained by putting the above solution(\ref{scalarfieldsolution}) back into the scalar
 field action (\ref{scalaraction}) and integrating over the extra dimension. This yields an 
 effective modulus potential at the visible brane as,
 \begin{eqnarray}
  V_{eff}=[2A^2k(2+\nu)\exp{(2\nu k\pi r)} + 2B^2k(\nu-2)]\nonumber\\
  -[ 2A^2\omega^2k\exp{(2\nu k\pi r)}- 2B^2\omega^2k(\nu-2)]\nonumber\\
  -[4A^2\omega^2k(2+\nu)/(1+\nu)\exp{((2+2\nu) k\pi r)}\nonumber\\
  - 4B^2\omega^2k(2-\nu)/(1-\nu)]\nonumber\\
  - [2AB\omega^2k(4-\nu^2)/(1-\nu^2)\exp{(2 k\pi r)}]
  \label{effectivepotential}
 \end{eqnarray}
 The boundary conditions given by eqn(\ref{boundarycondition1}) and
 eqn.(\ref{boundarycondition2}) yield the arbitrary constants A and B in the following form,
 \begin{eqnarray}
  A&=&[v_v\exp {-((2+\nu)k\pi r)}-v_h\exp{(-2\nu k\pi r)}]
  +\omega^2/2[v_v(2+\nu)/(1+\nu)\exp{-\nu k\pi r)}\nonumber\\
  &&-v_v(2-\nu)/(1-\nu)\exp{-3\nu k\pi r)}
  +2v_v(\nu/1-\nu^2)\exp{-((2+3\nu)k\pi r)}\nonumber\\
 &&+2v_h(\nu/1-\nu^2)\exp{-((2\nu-2)k\pi r)}]
  \label{arbitraryconstant1}
  \end{eqnarray}
  and
  \begin{equation}
  B=v_h[1+\omega^2/2(2-\nu)/(1-\nu)]-A[1+\omega^2(\nu/1-\nu^2)]
  \label{arbitraryconstant2}
 \end{equation}
 Putting A and B in expression (\ref{effectivepotential}) and minimizing the modulus potential, one gets the condition
 \begin{eqnarray}
  [v_v^2-v_h^2\exp{(-2\epsilon k\pi r)}]
  &+&\omega^2/2\exp{(2k\pi r)}[(v_v-v_h\exp{-(\epsilon k\pi r)})^2-8v_v^2\exp{-((6+\epsilon) k\pi r)}]\nonumber\\
  &-&\omega^2[v_v-v_h\exp{(-\epsilon k\pi r)}]=0
  \label{minimumpotentialcondition}
 \end{eqnarray}
 where we use $\nu=2+\epsilon$ with  $\epsilon=m^2/(4k^2)$ and ignore terms proportional to $\epsilon$.
 In this approximation eqn.(\ref{minimumpotentialcondition}) becomes
 \begin{equation}
  kr\pi=4(k^2/m^2)\ln{(v_h/v_v)}+(16/3)\omega^2(k^2/m^2)(v_v/v_h)^{(2+4/\epsilon)}
  \label{stabilisedmodulus1}
 \end{equation}
 Here $r$ is the stabilised distance between the two branes. If we now require that the same $r$ resolves the 
 gauge hierarchy problem as well, then the following condition holds,
 \begin{equation}
  \omega^2=[100\ln{(v_h/v_v)}-16\ln{(10)}]/[(1/4)10^{32}-(400/3)(v_v/v_h)^{402}]
  \label{finalresultvevratio1}
 \end{equation}
where we take $m/k\simeq0.2$. 
This result reveals that the ratio of the vev of the scalar field at the two branes depend on the brane cosmological constant. 
From the above relation (\ref{finalresultvevratio1}) between brane cosmological constant 
and vev ratio, we obtain Figure 1 ($\omega^2=-\varOmega/(3k^2)$) as, 
\begin{figure}[!h]
\begin{center}
 \centering
 \includegraphics[width=3.5in,height=1.70in]{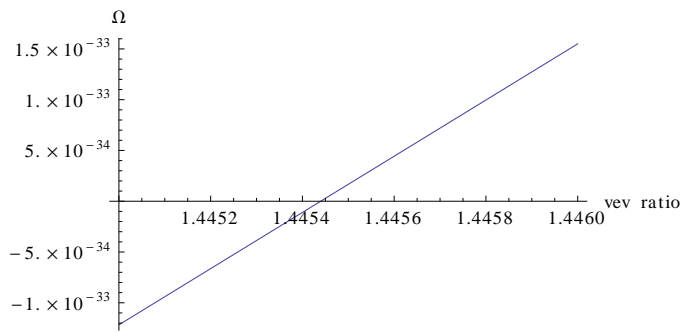}
 \caption{$(v_h/v_v)$ vs $\varOmega$}
 \label{plot omega vs vevratio for adsbrane}
\end{center}
\end{figure}

Figure 1 demonstrates that for a wide range of values of the brane cosmological constant, the vev ratio varies 
insignificantly and does not lead to any hierarchical values between the vevs .
\subsubsection{de-Sitter Brane ($\varOmega>0$)}
For de-Sitter brane we split the parameter space of cosmological constant into different regimes as following.
\begin{itemize}
 \item $0 \leq \varOmega \leq 10^{-32}$: \\
 Using the dS warp factor (\ref{dSwarpfactor}) one gets the scalar field solution for this regime as,
 \begin{eqnarray}
 \Phi(y)&=&[A\exp{((2+\nu) ky)} + B\exp{((2-\nu) ky)}] + \omega^2/2[A(2+\nu)/(1+\nu)\exp{((4+\nu) ky)}\nonumber\\
  &&+ B(2-\nu)/(1-\nu)\exp{((4-\nu) ky)}]
  \nonumber\\
 \end{eqnarray}
 Now proceeding similarly as in the AdS case, one ends up with the relation between brane cosmological constant and vev ratio as,
 \begin{equation}
 \omega^2=[16\ln{(10)-100\ln{(v_h/v_v)}}]/[(1/4)10^{32}-(400/3)(v_v/v_h)^{402}]
 \label{finalresultvevratio2}
\end{equation}
This leads to figure 2 ($\omega^2=\varOmega/(3k^2)$).
\begin{figure}[!h]
\begin{center}
 \centering
 \includegraphics[width=4.0in,height=1.70in]{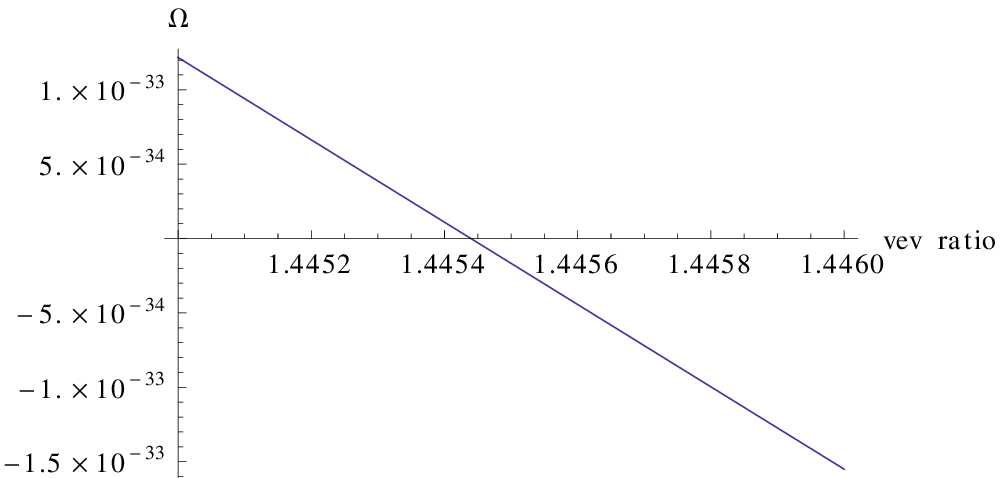}
 \caption{$(v_h/v_v)$ vs $\varOmega$}
 \label{plot omega vs vevratio for dsbrane1}
\end{center}
\end{figure}
\item {$10^{-32}\leq\varOmega\leq1$}:\\
 Using the dS warp factor (\ref{dSwarpfactor}), the scalar field solution for this regime is,
\begin{eqnarray}
 \Phi(y)=A\exp{-((\nu-2)ky)}_2F_1(2,2-\nu;1-\nu;(\omega^2/4)\exp{(2ky)}\nonumber\\
 +B\exp{((\nu+2)ky)}_2F_1(2,2+\nu;1+\nu;(\omega^2/4)\exp{(2ky)}
 \nonumber
\end{eqnarray}
where $_2F_1(arg)$ is the hypergeometric function. Keeping terms up to $\omega^2$, the above solution of scalar field becomes,
\begin{eqnarray}
 \Phi(y)&=&[A\exp{((2+\nu) ky)} + B\exp{((2-\nu) ky)}]  - \omega^2/2[A(2+\nu)/(1+\nu)\exp{((4+\nu) ky)}\nonumber\\
  &&+ B(2-\nu)/(1-\nu)\exp{((4-\nu) ky)}]
  \nonumber
\end{eqnarray}
Again proceeding similarly as before, one ends up with the relation between brane cosmological constant and vev ratio as,
\begin{equation}
 \omega^2=4(v_v/v_h)^{200}
 \label{finalresultvevratio3}
\end{equation}
This leads to  figure 3 ($\omega^2=\varOmega/(3k^2)$).
\begin{figure}[!h]
\begin{center}
 \centering
 \includegraphics[width=3.5in,height=1.70in]{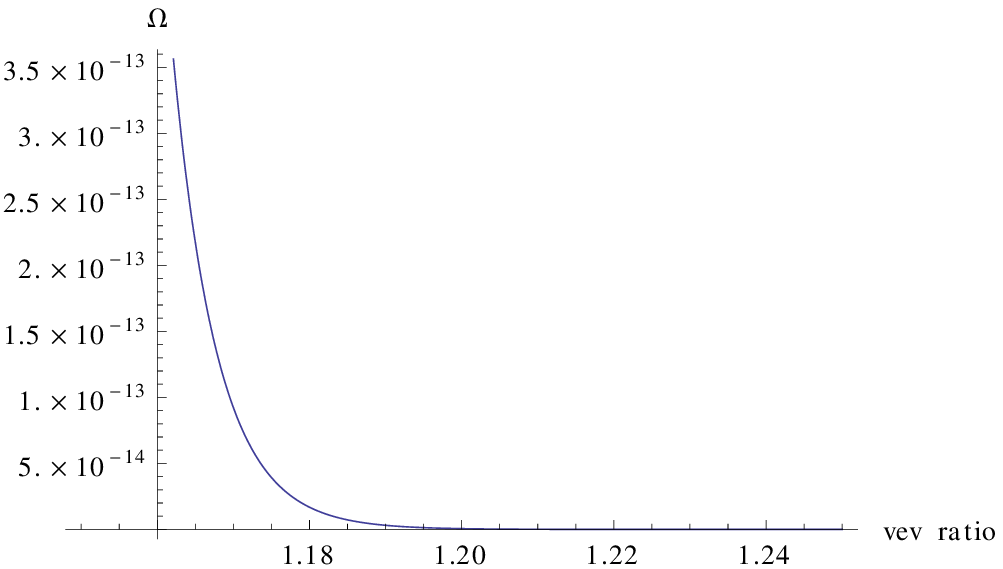}
 \caption{$(v_h/v_v)$ vs $\varOmega$}
 \label{plot omega vs vevratio for dsbrane2}
\end{center}
\end{figure}
\end{itemize}
Both Figures 2 and Figure 3 reveals that just as in Ads case, here also the deviation of the 
vev ratio of the scalar field from that of the GW-value is insignificant even when the brane is endowed with a large positive cosmological constant.\\
\subsection{Graviton Modes}
To study the graviton modes, we 
decompose the four-dimensional components of the metric into its Kaluza-Klein (KK) modes as,
\begin{eqnarray}
 g_{\mu \nu} (x,y)=\sum {h^n_{\mu \nu} (x)\xi_n(y)}
 \label{kaluzakleinmodes}
\end{eqnarray}
Here $h^n_{\mu\nu}(x)$ is the $n$ th KK graviton mode. Plugging back the decomposition in the action  and using the appropriate gauge conditions for  $h_{\mu\nu}(x)$\cite{csaki}, one gets 
\begin{equation}
 S_5=-(1/4)\int d^4xdy \sqrt{g} [\partial_M{g^{ij}}][\partial^M{g_{ij}}]
 \label{modifiedaction}
\end{equation}

This modified action leads to the equation of motion
\begin{equation}
 \eta ^{\mu \nu} \partial_\mu \partial_\nu h_{ij}^n(x)=m_n^2 h_{ij}^n(x)
 \label{equationforh}
\end{equation}
The above differential equation for $x$ dependent part of the metric holds if 
\begin{equation}
 \partial_y [\exp{(-4A(y))} \partial_y\xi^n]=-m_n^2 \exp{(-2A(y))} \xi^n(y)
 \label{equationforxi}  
\end{equation}
and the orthogonality condition 
\begin{equation}
 (1/r)\int \chi^n(y)\chi^m(y) dy = \delta_{mn}
 \nonumber
\end{equation}
are simultaneously satisfied. 

$m_n$ denotes the mass of nth KK graviton mode. 
Solution of eqn.(\ref{equationforxi}) as well as graviton KK modes for AdS and dS branes are now discussed in the following sections.
\subsubsection{Anti de-Sitter brane ($\varOmega<0$)}
As mentioned before that the range of cosmological constant on AdS brane is $-10^{-32}< \varOmega <0$.
Once again keeping terms up to $\omega^2$ and using the perturbative expansion of $\xi(y)$ and $m_n$ as,
\begin{eqnarray}
 \xi(y)&=&\xi_n^0(y)+\omega^2\xi_n^1(y)
 \nonumber\\
 m_n&=&m_n^0+\omega^2m_n^1
 \nonumber\\
\end{eqnarray}
(where $m_{n}^0$ is the mass of nth KK graviton mode on flat branes which is the RS scenario),
 eqn.(\ref{equationforxi}) becomes,
\begin{eqnarray}
 &[\partial_y(\exp{(-4ky)}\xi_n^0(y))+\exp{(-2ky)}(m_n^0)^2\xi_n^0]
 -\omega^2[\partial_y(\exp{(-4ky)}\xi_n^1(y))-\partial_y(\exp{(-2ky)}\xi_n^0(y))\nonumber\\
 &\exp{(-2ky)}(m_n^0)^2\xi_n^1-(1/2)(m_n^0)^2\xi_n^0+2m_n^0m_n^1\exp{(-2ky)}\xi_n^0]=0
 \label{finalequationforxi}
\end{eqnarray}
Defining the new variable $z_n^0=(m_n^0/k)\exp{(ky)}$, one gets the solution of $\xi_n(y)$ as,
\begin{equation}
 \xi_n(y)=(1/N_n)\exp{(2ky)}[J_2(z_n^0)+a_nY_2(z_n^0)-\omega^2f_n(z_n^0)]
 \label{solutionforxi}
 \end{equation}
 Now demanding the continuity of the graviton wave function at two branes, we get the 
 following boundary conditions
 \begin{eqnarray}
  \xi_n'(y=0)=0
  \nonumber\\
  \xi_n'(y=\pi r)=0
  \nonumber
 \end{eqnarray}
 From the first boundary condition and using the approximation $(m_n/k)\ll1$ one can conclude that
 the co-efficient $a_n$ is negligible. Thus the solution (\ref{solutionforxi}) becomes
 $\xi_n(y)=(1/N_n)\exp{(2ky)}[J_2(z_n^0)-\omega^2f_n(z_n^0)]$. The other boundary condition yields,
 \begin{eqnarray}
  &\exp{(k\pi r)}(m_n^0/k)J_1(\exp{(k\pi r)}(m_n^0/k))-2\omega^2kf_n(\exp{(k\pi r)}(m_n^0/k))\nonumber\\
  &-\omega^2\exp{(k\pi r)}(m_n^0/k)f_n'(\exp{(k\pi r)}(m_n^0/k))=0
  \label{finalresultgravitonmode1}
 \end{eqnarray}
 which leads to first order correction of graviton KK mass modes as,
 \begin{eqnarray}
  m_{(n=1)}&=&m^0_{n=1}+3.5*10^{30}*\omega^2
  \nonumber\\
  m_{(n=2)}&=&m^0_{n=2}+2*10^{30}*\omega^2
  \nonumber
 \end{eqnarray}
 \subsubsection{de-Sitter Brane($\varOmega >0$)}
 As before for 
 \begin{itemize}
 \item $10^{-32} \leq \varOmega \leq 1$ using 
 eqn. (\ref{dSwarpfactor}), expression for warp factor becomes,
\begin{equation}
 \exp{[-4A(y)]}=\exp{(-4ky)}[1-\omega^2\exp{(2ky)}]
 \nonumber
\end{equation}
Taking this expression of warp factor and proceeding similarly, one ends up with following graviton mass correction due to 
brane cosmological constant.
 \begin{eqnarray}
   m_{(n=1)}&=&m^0_{n=1}+(2.5*10^{-10})
  \nonumber\\
  m_{(n=2)}&=&m^0_{n=2}+(13.44*10^{-10})
  \nonumber
 \end{eqnarray}
 for $\varOmega \sim 10^{-20}$
 \end{itemize}
 In a similar way we can extend our analysis for very large values of brane cosmological 
 constant ($\Omega$). \\
 We now summarise our results for different cases in the following table (for $k/M_{pl}\sim0.1$).
 \begin{table}[!h]
  \centering
  \begin{tabular}{|c| c| c| c|c|}
   \hline \hline
   $\varOmega$ & $kr$ & $v_h/v_v$ & $m_{(n=1)}=(m_{n=1}^0+\Delta m_{n}) (TeV)$ & $m_{(n=2)}=(m_{n=2}^0+\Delta m_{n}) (TeV)$\\
   \hline
   $-10^{-32}$ & $\sim11$ & 1.446 & (0.383+0.03) & (0.702+0.02)\\
   0 & $\sim11$ & 1.445439771 & 0.383 & 0.702\\
   $10^{-20}$ & $\sim10$ & 1.3700 & (0.383+0.44) & (0.702+0.55)\\
   100 & 0.095 & $\sim1$ & 10.5263 & 21.0526\\
   625 & 0.039 & $\sim1$ & 25.641 & 51.282\\
   $10^{4}$ & 0.0099 &  $\sim1$ & 101.01 & 202.02\\
   \hline
  \end{tabular}
  \caption{Warping parameter ($kr$), vev ratio ($v_h/v_v$) and graviton mass modes ($m_n$) for a wide range of
  cosmological constant ($\varOmega$)}
  \label{Table-3}
 \end{table}

From the above table, it is  evident that the ratio of $v_h$ and $v_v$ is of the order of unity for the entire chosen range 
 of values of brane cosmological constant. This condition justifies the fact that the back-reaction of the stabilising scalar field on background 
 spacetime can be neglected even in the presence of brane cosmological constant. Again from \cite{ssg}, it turns out that the perturbation 
 of visible brane tension due to brane cosmological constant is given by,
 \begin{equation}
  \delta V_v = 12M^3k [\omega^2\exp{(2kr\pi)}]/[1+\sqrt{(1-\omega^2)}+(\omega^2/2)\exp{(2kr\pi)}]
  \label{perturbation}
 \end{equation}
 Since for AdS brane,  $\omega^2=-\varOmega/(3k^2)$ therefore  from  eqn.(\ref{perturbation}), it is easy to see  that $\delta V_v<kv_v^2$. 
 This immediately  ensures \cite{GW} that the minimum is a global one. Similar argument also holds for dS brane. 

\section{Conclusion}
 We now summarize the findings and the implications of our results.
 \begin{itemize}
   \item We have demonstrated that the extra dimensional modulus can be stabilised by Goldberger-Wise mechanism for a wide range of values of cosmological constant 
  both in de-Sitter and anti de-Sitter region.
  It has been shown in  \cite{GW} that if the vev ratio of the scalar field in the bulk
  is of the order $\sim 1.46$ or less than that, then one can safely ignore the back reaction 
  of the scalar field on background spacetime for the purpose of modulus stabilisation. Now from our above table it 
  is evident that since the vev ratio lies between 
  $1<(v_h/v_v)<1.46$ for the entire parameter space of cosmological constant, 
  therefore the back reaction can be sagely ignored even in the ``Generalised Randall Sundrum Scenario''.
  In this sense the Goldberger-Wise stabilisation mechanism is extremely robust against the extent of non-flatness of our universe. Our result also reveals that even for non-flat branes the modulus potential continues to yield  a global minimum ensuring a robust modulus stabilisation against   
 perturbations. \\
   \item We have derived
  the modifications of the KK graviton mass modes due to the presence of a non-zero cosmological constant on the brane in the  
  generalised Randall Sundrum scenario. We found that the masses of the graviton KK modes increases with brane cosmological constant and 
  may deviate significantly from the values estimated in RS scenario as the values of the brane cosmological constant increases.
  During this analysis, we restricted the choice of the parameters in a region so that the gauge hierachy problem can simultaneously be  resolved.  
  In the context of the present epoch of our universe ( visible 3-brane ) , these results indicate that due to extreme smallness of the 
  value of the cosmological constant  ($10^{-124}$ in Planckian unit), the warped model resembles very closely to the RS model with graviton 
  KK mode masses $\sim TeV$. However this scenario will change
  significantly in epoch with a large cosmological constant.  
  \end{itemize}
  
  \section{Acknowledgements}
  We thank A. Das for illuminating discussions.


\begin{thebibliography}{90}
   \bibitem{RS}
L. Randall and R. Sundrum, Phys. Rev. Lett. 83 (1999) 4922.
\bibitem{GW}
W.D. Goldberger and M.B. Wise, Modulus stabilization with bulk fields, Phys. Rev.
Lett. 83 (1999) 4922 [hep-ph/9907447]
\bibitem{atlas1}
ATLAS Collaboration, Phys.Lett.B710 (2012) 538-556
\bibitem{atlas2}
ATLAS Collaboration, G. Aad et al, Phys.Rev.D.90, 052005 (2014)
\bibitem{dhr}
H. Davoudiasl, J.L. Hewett, T.G. Rizzo,Phys.Rev. Lett. 84(2000)2080
\bibitem{ashmita}
A. Das and S. Sengupta, arxiv:1506.05613[hep-ph] 
\bibitem{rizzo}
T. G. Rizzo, Int.J.Mod.Phys A15 (2000) 2405-2414
\bibitem{ashmita1}
A. Das and S. Sengupta, arxiv:1303.2512[hep-ph]
\bibitem{yong}
Y. Tang, JHEP 1208 (2012) 078
\bibitem{dhr1}
H. Davoudiasl, J.L. Hewett, T.G. Rizzo, JHEP 0304 (2003) 001
\bibitem{thomas}
M. T. Arun, D. Choudhury, A. Das, S, Sengupta, Phys.Lett.B746 (2015) 266-275
\bibitem{ssg}
S. Das, J. Maity and S. Sengupta, JHEP,0805 (2008) 042
\bibitem{joydip}
J. Mitra, S. Sengupta : Phys.Lett.B683 (2010) 42-49
\bibitem{koley}
R. Koley, J. Mitra, S. Sengupta ; Europhys.Lett.85 (2009) 41001
\bibitem{rubakov}
V.A. Rubakov and M.E. Shaposhnikov, Do we live inside a domain wall? , Phys.Lett. B125,136(1983)
\bibitem{csaki}
C. Csaki; Tasi lectures on extra dimensions and branes, arXiv:hep-ph/0404096v1 
 \end{thebibliography}
\end{document}